\documentclass[12pt,runningheads,a4paper]{amsart}
\usepackage{amssymb}
\usepackage{amsmath}

\begin{document}
\title{ Kinetic Equations in the theory of Normal Fermi Liquid}
\author[ A. S. Kondratyev${}^{1,2}$, I. Siddique${}^1$]
{A. S. Kondratyev${}^{1,2}$, I. Siddique${}^1$}
\address{${}^1$School of Mathematical Sciences, GC University\\
68-B, New Muslim Town, Lahore, Pakistan. e-mail: \
imransmsrazi@gmail.com
\\${}^2$Department of Physics, Herzen State Pedagogical
University\\Moika River Embankment, 48 191168 St. Petersburg,
Russia. e-mail: kondrat98926@yahoo.com}
 \maketitle

\begin{abstract}
On the bases of the improved approximation for the spectral function
of one-particle states the Landau-Silin kinetic equations for the
normal Fermi liquids of neutral and electrically charged particles
are shown to be valid at finite temperature above the temperature of
superfluid transition.
\end{abstract}

\section{Introduction}
\fontsize{12}{24}\selectfont

The exclusive successfulness of the phenomenological Landau normal
Fermi liquid theory \cite{1} in predicting and describing a set of
new phenomena, among them the zero sound in liquid He$^3$ and spin
waves in nonferromagnetic metals, made this theory a subject of
investigation on the basis of strict microscopic theory. Most
attention was devoted to the derivation of the kinetic equation for
the quasiparticle distribution function. The initial derivation of
this equation was produced by Kadanoff and Byam \cite{2} on the
basis of the quasiparticle (QP) approximation for the spectral
function and was continued by some followers who used so called
extended quasiparticle (EQP) approximation \cite{3}--\cite{7}.
However, in all these cases the second Poisson bracket in the left
side of Kadanoff-Baym (KB) generalized kinetic equation could not be
eliminated in a lawful mathematical way (see below). This fact made
the temperature range of validity of the kinetic equation for
quasiparticle distribution very narrow, strictly speaking the theory
was proved to be valid only in the vicinity of absolute zero.

Experimental discovery of the superfluidity of He$^3$ at the
temperature lower than the temperature at which the zero-sound in
the normal Fermi liquid was discovered, and theoretical works
devoted to the description of the superfluid state \cite{8,9}, left
no room for the temperature range of validity of the equation for
normal Fermi liquid. At the same time theoretical predictions of
this theory turned to be in a perfect numerical agreement with the
experimental data. This fact stimulated a second wave of attempts to
deriving the kinetic equation for normal Fermi liquid, but as it was
mentioned above, the result was not achieved to a satisfactory
extent. Further development of KB theory went in the direction of
the constructing of a nonlocal quasiparticle kinetic equations
\cite{10,11}, a development of a detailed selfconsistent microscopic
treatment of arbitrary initial correlations in the system \cite{12},
etc., but the question of the temperature range of validity of the
Landau's kinetic equation remained open.

\section{Spectral function and kinetic equation for normal Fermi
 Liquid}

The problem consisted in a mathematically lawful elimination of the
second generalized Poisson bracket (a"puzzling term" in the
terminology accepted in \cite{5}) in the generalized Kadanoff-Byam
kinetic equation written in the collisionless approximation in the
case of slowly varying in space and time disturbances when only the
first derivatives with respect to $T$ and $\vec{R}$ are taken into
account \cite{2}:
\begin{equation}
[\omega-e(\vec{p}\omega;\vec{R}T),\text
g^<(\vec{p}\omega;\vec{R}T)]+ [\text Re \ \text
g(\vec{p}\omega;\vec{R}T),\sigma^<(\vec{p}\omega;\vec{R}T)]=0,
\end{equation}
here [A,B]-the generalized Poisson bracket, defined by the
expression:
\begin{equation}
[A,B]=\frac{\partial{A}}{\partial\omega}\frac{\partial{B}}{\partial{T}}-
\frac{\partial{A}}{\partial{T}}\frac{\partial{B}}{\partial\omega}
-\nabla_{\vec{p}}A\ .\ \nabla_{\vec{R}}B+\nabla_{\vec{R}}A\ .\
\nabla_{\vec{p}}B,
\end{equation}
$e(\vec{p}\omega;\vec{R}T)$-the energy of a particle, defined by the
equality:
\begin{equation}
e(\vec{p}\omega;\vec{R}T)=E^{HF}(\vec{p};\vec{R}T)+\text Re\
\sigma(\vec{p}\omega;\vec{R}T),
\end{equation}
$E^{HF}$ is a one-particle energy in the Hartree-Fock approximation,
$\text Re\ \sigma$ is a real part of the correlation energy of a
particle related to the imaginary part through the Hilbert
transform,
\begin{equation}
\text Re\ \sigma (\vec{p}\omega ;\vec{R}T)=P \int_{-\infty
}^{\infty}\frac{d{\omega'}}{2\pi}
\frac{\Gamma{(\vec{p}}{\omega'};\vec{R}T)}{\omega-\omega'}.
\end{equation}
Here, P refers to a principle value integration.\\It was shown in
\cite{2} that the function $\text g$ entering the Eq. (1) can be
taken in the form
\begin{equation}
\text g(\vec{p}Z;\vec{R}T)=[Z-E^{HF}(\vec{p};\vec{R}T)-\text Re\
\sigma(\vec{p}Z;\vec{R}T)]^{-1}.
\end{equation}
Eqs. (4) and (5) lead to the following general form for a spectral
function of one-particle states in the system:
\begin{equation}
a(\vec{p}\omega;\vec{R}T)=\frac{\Gamma(\vec{p}\omega;\vec{R}T)}
{[\omega-e(\vec{p}\omega;\vec{R}T)]^2+\frac{\Gamma^2(\vec{p}\omega;\vec{R}T)}{4}}.
\end{equation}
The spectral function satisfies the exact sum rule:
\begin{equation}
\int_{-\infty
}^{\infty}\frac{d{\omega}}{2\pi}a(\vec{p}\omega;\vec{R}T)=1.
\end{equation}
Correlation function $\text g^<$ is related to a spectral function
by the equality [2]:
\begin{equation}
\text
g^<(\vec{p}\omega;\vec{R}T)=a(\vec{p}\omega;\vec{R}T)f(\vec{p}\omega;\vec{R}T).
\end{equation}
The complexity of the expression (6) makes to search for a certain
approximations for the spectral function which could be successfully
used in calculations.

Quasiparticle approximation for the spectral function (6)
corresponds to the case $\Gamma\rightarrow{0}$, when the
quasiparticles are stable:
\begin{equation}
a_{QP}=2\pi{Z(\vec{p};\vec{R}T)}\delta[\omega-E(\vec{p};\vec{R}T)],
\end{equation}
where $E(\vec{p};\vec{R}T)$ is the solution of the equation:
\begin{equation}
E(\vec{p};\vec{R}T)=E^{HF}(\vec{p};\vec{R}T)+\text Re\
\sigma(\vec{p}\omega; \vec{R}T)\vert_{\omega=E(\vec{p};\vec{R}T)},
\end{equation}
and the renormalizing factor $Z(\vec{p};\vec{R}T)$ is given by the
expression:
\begin{equation}
Z^{-1}(\vec{p};\vec{R}T)=1-\frac{\partial\ Re\
\sigma(\vec{p}\omega;\vec{R}T)}{\partial\omega}
\bigg|_{\omega=E(\vec{p};\vec{R}T)}.
\end{equation}
Using the approximation (9) one follows the KB way to deriving  the
Landau's kinetic equation when the second Poisson bracket in Eq. (1)
is dropped considering $\sigma^<=f\Gamma$ to be negligibly small
what is valid only at $\omega=\mu$ at zero temperature. \\The
extended quasiparticle (EQP) \cite{3}--\cite{7} and improved
extended quasiparticle (iEQP) \cite{4} approximations for the
spectral function are written in the form:
\begin{equation}
a_{EQP}=2\pi{Z\delta(\omega-E)}+P\frac{\Gamma}{(\omega-E)^2},
\end{equation}
\begin{equation}
a_{iEQP}=2\pi{Z\delta(\omega-E)}+ZP\frac{\Gamma}{(\omega-E)^2},
\end{equation}
where $E$ and $Z$ are determined by the Eqs. (10) and (11)
correspondingly.

It is easy to verify that the approximations (12) and (13) do not
satisfy the KB equation for a nonequilibrium spectral function:
\begin{equation}
[\omega-e(\vec{p}\omega;\vec{R}T),a(\vec{p}\omega;\vec{R}T)]+
[\text{Re}\ \text
g(\vec{p}\omega;\vec{R}T),\Gamma(\vec{p}\omega;\vec{R}T)]=0,
\end{equation}
and do not lead to the elimination of the second Poisson bracket in
Eq. (1) in the case of finite $\Gamma$. Only such a mathematically
lawful elimination of this term for finite values of $\Gamma$ would
testify the validity of the Landau's equation at a finite range of
temperature.

The expressions (12) and (13) both were obtained on  the basis of
the general expression (6) by means of the Taylor expansion in
powers of $\Gamma$ in the frame of different approximations
\cite{3}--\cite{7}. As a result, some important factors were lost.
In reality there does not exist a mathematically strict correct form
for the expansion of (6) in power series of $\Gamma$ which starts
with the delta function when $\Gamma=0$.

Another expression for the spectral function can be offered on the
basis of the following consideration. We utilize a well known
relation of the Fourier transform in the case of a constant value of
c:
\begin{equation}
\int_{-\infty}^{\infty}e^{-|t|c}e^{\imath{tx}}dt=\frac{2c}{c^2+x^2}\hspace{.1cm}
 ,\hspace{.2cm} c>0.
\end{equation}
Expanding the first exponent in the left side of (15) in Taylor
series, we get:
\begin{equation}
\frac{2c}{c^2+x^2}=\int_{-\infty}^{\infty}(1-c|t|+\frac{{c^2t^2}}{2!}-...)e^{\imath{tx}}dt.
\end{equation}
Now we use the formulas equivalent to those represented in [13,14]:
\begin{eqnarray}
\int_{-\infty}^{\infty}t^{(2n)}e^{\imath{tx}}dt&=&
2\pi (-\imath)^{2n}\delta^{(2n)}(x) ,\hspace{.2cm} n=0,1,2,...\ , \\
\int_{-\infty}^{\infty}|t|^{(2n+1)}e^{\imath{tx}}dt &=&
-2\sin((2n+1)\frac{\pi}{2})(2n+1)!\frac{1}{|x|^{2n+2}},\\\nonumber\hspace{.1cm}n=0,1,2,...\
.
\end{eqnarray}

If the quantity $\Gamma$ in Eq. (6) was not a function of $\omega$,
then the expressions (17), (18) would lead to a strict correct
expansion of the spectral function (6) in terms of the power series
of $\Gamma$. In the case of $\Gamma$ depending on $\omega$, one can
rely only on the first two terms of the expansion: the delta
function independent of $\Gamma$ and the term proportional to
$\Gamma$. Then, taking into account that
\begin{equation}
\omega-e(\vec{p}\omega;\vec{R}T)=Z^{-1}(\vec{p};\vec{R}T)[\omega-E(\vec{p};\vec{R}T)],
\end{equation}
we come to the following approximation for the spectral function
(6):
\begin{equation}
a=2\pi Z\delta(\omega-E)+Z^2 P\frac{\Gamma}{(\omega-E)^2}.
\end{equation}

It is not difficult to verify that the approximation (20) satisfies
the sum rule (7), with the same precision as approximations (12) and
(13) do, but contrary to them the approximation (20) satisfies the
Eq. (14) for the spectral function and eliminates the second Poisson
bracket in the left side of Eq. (1). Indeed, when we substitute Eq.
(20) into Eq. (14), the first term in the right of this expression
gives:
\begin{equation}
[\omega-e, 2\pi Z\delta(\omega-E)]= 2\pi Z[(\omega-E)Z^{-1},
\delta(\omega-E)]=0,
\end{equation}
due to the property of the generalized Poisson bracket
\\$$[A,f(A)]=0.$$\\
The second term in the right side of Eq. (20) leads to the
expression:
\begin{equation}
\Big[\omega-e,Z^2\frac{\Gamma}{(\omega-E)^2}\Big]=Z
\frac{1}{(\omega-E)^2}[\omega-E,\Gamma],
\end{equation}
the second poisson bracket in Eq. (14) due to Eq. (5) gives the
expression:
\begin{eqnarray}
[\text{Re}\ \text g,\Gamma]=
Z\Big[\frac{1}{(\omega-E)},\Gamma\Big]&=
-Z\frac{1}{(\omega-E)^2}[\omega-E,\Gamma].&
\end{eqnarray}
Thus, the $ansatz$ (20) satisfies the Eq. (14) exactly. Now we
consider the Eq. (1). Using Eq. (8), we get:
\begin{equation}
[\omega-e,af]+[\text{Re}\ \text g,f\Gamma]=0.
\end{equation}

The substitution of the first term in Eq. (20) into Eq. (24) leads
directly to the kinetic equation for the quasiparticle distribution
function $n(\vec{p};\vec{R}T)$ \cite{2}:
\begin{equation}
\frac{\partial n}{\partial T}+\nabla_{\vec{P}}E\ .\
\nabla_{\vec{R}}n-\nabla_{\vec{R}}E\ .\ \nabla_{\vec{p}}n=0,
\end{equation}
$$n(\vec{p};\vec{R}T)=f(\vec{p}\omega;\vec{R}T)\mid_{\omega=E(\vec{p};\vec{R}T)}.$$
The second term in Eq. (20), being substituted to Eq. (24), gives
the expression:
\begin{equation}
Z^2\Big[\omega-e,\frac{\Gamma f}{(\omega-E)^2}\Big]=Z
[\omega-E,\Gamma f]\frac{1}{(\omega-E)^2},
\end{equation}
which is eliminated by the second poisson bracket in Eq. (24):
\begin{eqnarray}
\nonumber[\text{Re}\ \text g,\sigma^<]=[\frac{1}{\omega-e},\Gamma
f]=
Z\Big[\frac{1}{\omega-E},\Gamma f\Big]&= \\
Z[\Gamma f,\omega-E]\frac{1}{(\omega-E)^2}.&
\end{eqnarray}
Thus, the kinetic Eq. (25) is valid for finite values of $\Gamma$
and correspondingly, for nonzero temperatures. For qualitative
estimation of the precision, we can use the third term in the
expansion (16) which is proportional to $\Gamma^2$. Substituting
this term into Eq. (24), it is not difficult to show that this
equation is valid up to the terms of order $\Gamma^2$.

\section{ kinetic equation for normal Fermi liquid in a magnetic field}

The kinetic equation for the normal Fermi liquid consisting of
charged particles in the presence of compensating background was
considered in [15] in the quasiparticle approximation for the
spectral function. It was shown that the spin splitting of the
energy levels being neglected, the spectral function for a system in
a nonquantizing magnetic field can be written in the form:
\begin{equation}
a(\vec{p}\omega;\vec{R}T)=\frac{\Gamma(\vec{p}\omega;\vec{R}T)}
{[\omega-\frac{(\vec{p}-\frac{e}{c}\vec{A}(\vec{R}T))^2}{2m}-u(\vec{R}T)-\sigma^{HF}
(\vec{p};\vec{R}T)-\text{Re}\ \sigma(\vec{p}\omega;\vec{R}T)]^2
+\frac{\Gamma^2(\vec{p}\omega;\vec{R}T)}{4}},
\end{equation}
where $\vec{p}$ is the canonical momentum and $u(\vec{R}T)$,
$\vec{A}(\vec{R}T)$ are scalar and vector potentials of the
electromagnetic field, $\vec{A}(\vec{R}T)$ being chosen in the
Coloumb gauge $div\vec{A}(\vec{R}T)$=0. The kinetic equation for
quasiparticle distribution is written in the form (25) \cite{15} and
acquires a usual form after the transition to kinetic momentum
$\vec{P}=\vec{p}-(e/c)\vec{A}(\vec{R}T)$ \cite{1}. We would like to
stress that the kinetic equations for normal Fermi liquid were
written in \cite{16} right in the form, corresponding the expression
(28) for the spectral function, concerning the dependence of all the
quantities on vector potential $\vec{A}$ and canonical momentum
$\vec{p}$ and then were transformed into gauge-invariant form
\cite{17}. More details about the development of transport theory of
interacting fermions in an electromagnetic field can be found in
\cite{18}--\cite{20}.

The derivation of the phenomenological Landau-Silin kinetic
equations in the case when spin splitting of energy levels is taken
into account was produced in \cite{21} also in the quasiparticle
approximation. In this case all quantities become matrices in spin
space, in particular the quantity $e(\vec{p}\omega;\vec{R}T)$ has
the form:
\begin{equation}
e(\vec{p}\omega;\vec{R}T)=\Big[\frac{(\vec{p}-\frac{e}{c}\vec{A}(\vec{R}T))^2}{2m}+
u(\vec{R}T)\Big]I-\frac{e}{mc}\mathop{\vec{S}}\limits^\wedge.curl\vec{A}(\vec{R}T)
+\tilde{\sigma}(\vec{p}\omega;\vec{R}T),
\end{equation}
where $I$ and $\mathop{\vec{S}}\limits^\wedge$ are the unit matrix
and the set of three spin matrices \cite{1}, $\tilde{\sigma}$ is a
Hermitian part of the self-energy matrix $\sigma$. The spectral
function is still given by Eq. (28). The spectral function $a$ and
the inverse life-time of a particle's state $\Gamma$ are Hermitian
matrices in the case under consideration.

When $\Gamma$ is considered to be finite (nonzero), the next
possibilities can occur. If the magnetic field is not strong, so
that $\Gamma$ exceeds the spin splitting of the energy levels, the
last is not essential and $e(\vec{p}\omega;\vec{R}T)$ can be written
in the form:
\begin{equation}
e(\vec{p}\omega;\vec{R}T)=\frac{(\vec{p}-\frac{e}{c}\vec{A}(\vec{R}T))^2}{2m}+
u(\vec{R}T)+\sigma^{HF}(\vec{p};\vec{R}T)+\text Re\
\sigma(\vec{p}\omega;\vec{R}T).
\end{equation}
The approximation for the spectral function in this case can be
written in a way analogous to Eq. (20):
\begin{equation}
a(\vec{p}\omega;\vec{R}T)=2\pi{Z(\vec{p};\vec{R}T)}
\delta(\omega-E(\vec{p};\vec{R}T))+Z^2P\frac{\Gamma}{(\omega-E)^2},
\end{equation}
where $E$ is a solution of Eq. (10) with $e$ given by Eq. (30).

The situation turns to be more complicated when spin splitting of
the energy levels should be taken into account. More convenient form
of the theory suitable for the generalization of the quasiparticle
approximation for the spectral function, than that presented in
\cite{21}, was developed in \cite{22}. First of all , we should
stress, that even in the quasiparticle approximation for the
spectral function the kinetic equation of the phenomenological
theory are valid only with the precision to the squared ratio of
spin splitting of the energy levels to the chemical potential of the
system \cite{22}. Thus, producing the derivation of the  kinetic
equation with the improved approximation for the spectral function
we should take into account only the terms that do not exceed this
precision.
\\We start with the expansion of the matrices $\text g^<$ and $e$ over the
full set of matrices in spin space,
\begin{equation}
\text g^<=\frac{1}{2}{\text
g_0}I+\mathop{\vec{S}}\limits^\wedge.\mathop{\vec{\text
g}}\limits^\wedge,
\end{equation}
\begin{equation}
e={e_1}I+2\mathop{\vec{S}}\limits^\wedge.\mathop{\vec{e_2}}\limits^\wedge.
\end{equation}

These expansions should be substituted into the generalized KB
kinetic equations which under consideration are written in the form
\cite{22}:
\begin{equation}
(\omega I-e+\frac{\iota}{2}\Gamma)\text g^<+\frac{\iota}{2}[\omega
I-e+\frac{\iota}{2}\Gamma,\text g^<]-\sigma^<(\tilde{\text
g}+\frac{\iota}{2}a) -\frac{\iota}{2}[\sigma^<,\tilde{\text
g}+\frac{\iota}{2}a]=0,
\end{equation}
\begin{equation}
\text g^<(\omega I-e-\frac{\iota}{2}\Gamma)+\frac{\iota}{2}[\text
g^<,\omega I-e-\frac{\iota}{2}\Gamma]-(\tilde{\text
g}-\frac{\iota}{2}a)\sigma^< -\frac{\iota}{2}[\tilde{\text
g}-\frac{\iota}{2}a,\sigma^<]=0,
\end{equation}
where $\tilde{\text g}$ is a Hermitian part of the matrix $\text
g(\vec{p}Z;\vec{R}T)$.\\ Separating in Eqs. (34) and (35) the terms
corresponding to the collisionless case, we get in the quasiparticle
approximation ($\Gamma$=0) the equations:
\begin{equation}
[\omega-e_1,\text g_0]-[e_{2i},\text g_i]=0,
\end{equation}
\begin{equation}
[\omega-e_1,\vec{\text g}]-[e_2,\text
g_0]-2\vec{e_2}\times\vec{\text g}=0,
\end{equation}
\begin{equation}
(\omega-e_1)\text g_0-\vec{e_2}.\vec{\text g}=0,
\end{equation}
\begin{equation}
(\omega-e_1)\text g_i-e_{2i}\text g_0=0.
\end{equation}
Here, the quantities with Latin indexes are the cartesian
projections of the corresponding vectors; summation over the
repeated indexes is supposed to be done. Eqs. (36)--(39) lead to the
kinetic equations of the phenomenological Landau-Silin theory
\cite{22} in the limit $\Gamma\rightarrow 0$: Eqs. (38) and (39)
give the expressions for the spectral functions, while the Eqs. (36)
and (37) give kinetic equations on the basis of the determined
spectral functions.\\With the help of Eqs. (38) and (39) we get the
following expressions for the functions $\text g_0$ and $\vec{\text
g}$ in the quasiparticle approximations \cite{22}:
\begin{equation}
\text
g_0=f_\uparrow\delta(\omega-e_\uparrow)+f_\downarrow\delta(\omega-e_\downarrow),
\end{equation}
\begin{equation}
\vec{\text g}=\vec{f_\uparrow}\delta(\omega-e_\uparrow)+
\vec{f_\downarrow}\delta(\omega-e_\downarrow)+\vec{f}\delta(\omega-e_1),
\end{equation}
where
\begin{equation}
e_\uparrow=e_1-|\vec{e_2}| \,\,\,;e_\downarrow=e_1+|\vec{e_2}|.
\end{equation}
Vector $\vec{f_\uparrow}$ is antiparallel to $\vec{e_2}$,
$\vec{f_\downarrow}$ is parallel to $\vec{e_2}$, and $\vec{f}$ is
perpendicular to $\vec{e_2}$.

The argument of all the delta functions in (40), (41) except one of
them, are not equal to $\omega-e_1$. Thus, the substitution of
expressions (40), (41) into the Eqs. (36) and (37) will generate the
terms with the derivatives of the $\delta-$ functions. As a result,
after rather tedious transformations of Eq. (36) we get the Eq.
(19.41) from [22]. The analysis of the distinction between different
renormalizing factors entering this equation, leads to the mentioned
above conclusion about the precision of the validity of the
phenomenological kinetic equation in the case $\Gamma=0$, i.e at
zero temperature. In the case of finite $\Gamma$, (but $\Gamma$
being less then the spin splitting) it is necessary to put down the
system of Eqs. (36)--(39) in the collisionless approximation, but
saving the terms with $\Gamma$, which do not enter the collision
integrals. It can be done, tracing carefully what terms with
$\Gamma$ in the equations in the absence of magnetic field do not
enter collision integrals. Thus, we come to the system of equations:
\begin{equation}
(\omega I-e)\text g^<+\frac{\iota}{2}[\omega I-e,\text g^<]
-\sigma^<\tilde{\text g}-\frac{\iota}{2}[\sigma^<,\tilde{\text
g}]=0,
\end{equation}
\begin{equation}
\text g^<(\omega I-e)+\frac{\iota}{2}[\text g^<,\omega I-e]
-\tilde{\text g}\sigma^<-\frac{\iota}{2}[\tilde{\text
g},\sigma^<]=0,
\end{equation}
where the expansions of the type (32), (33) should be done for the
functions $\tilde{g}$ and $\sigma^<=\Gamma f$:
\begin{equation}
\tilde{\text g}=\frac{1}{2}\tilde{\text
g_0}I+\mathop{\vec{S}}\limits^\wedge.\vec{\tilde{\text g}},
\end{equation}
\begin{equation}
\sigma^<=\sigma{_1}^<I+2\mathop{\vec{S}}\limits^\wedge.{\vec{\sigma_2}}^<.
\end{equation}
Finally, taking into account the comment about the terms, exceeding
the precision of the equation's validity, we get the system of
equations that will lead to the kinetic equations:
\begin{equation}
[\omega-e_1,\text g_0]-[e_{2i},\text g_i]+[\tilde{\text
g_0},\sigma_1^<]=0,
\end{equation}
\begin{equation}
[\omega-e_1,\vec{\text g}]-[\vec e_2,\text
g_0]-2\vec{e_2}\times\vec{\text g}+[\tilde{\text
g_0},{\vec{\sigma_2}}^<]=0,
\end{equation}
and the system of equations that will give the expressions for
spectral functions:
\begin{equation}
(\omega-e_1)\text g_0-\vec{e_2}.\vec{\text g}-\tilde{\text
g_0}\sigma_1^<=0,
\end{equation}
\begin{equation}
(\omega-e_1)\text g_i-e_{2i}\text g_0- \tilde{\text
g_0}\sigma_{2i}^<=0.
\end{equation}
Now we consider the case $e_\uparrow=e_1-|\vec{e_2}|$ and choose the
approximations for $\text g_0$ and $\vec{\text g}$ in the form:
\begin{equation}
\text g_0=Z_\uparrow f_\uparrow
\delta(\omega-E_\uparrow)+Z^2\frac{\sigma_1^<}{(\omega-E)^2},
\end{equation}
\begin{equation}
\vec{\text g}=Z_\uparrow \vec{f_\uparrow}
\delta(\omega-E_\uparrow)+Z^2\frac{\vec{\sigma_2}^<}{(\omega-E)^2},
\end{equation}
where $E_\uparrow $ and E are the solutions of the equations:
\begin{equation}
E(\vec{p};\vec{R}T)=e_1(\vec{p};\vec{R}T)|_{\omega=E},
\end{equation}
\begin{equation}
E_\uparrow(\vec{p};\vec{R}T)=[e_1(\vec{p}\omega;\vec{R}T)
-|e_2(\vec{p};\vec{R}T)|]|_{\omega=E},
\end{equation}
and renormalizing factors $Z_\uparrow$ and Z are defined by the
expressions:
\begin{equation}
Z_\uparrow^{-1}(\vec{p} ;\vec{R}T)=1-\frac{\partial\
e_\uparrow(\vec{p}\omega ;\vec{R}T)}{\partial \omega}
\bigg|_{\omega=E_\uparrow},
\end{equation}
\begin{equation}
Z^{-1}(\vec{p} ;\vec{R}T)=1-\frac{\partial\ e_1(\vec{p}
\omega;\vec{R}T)}{\partial\omega}\bigg|_{\omega=E}.
\end{equation}
If we substitute the expressions (51) and (52) into Eqs. (49) and
(50), these equations would be satisfied with the precision to the
terms of the order $\Gamma^2$, $|\vec{e_2}|^2$ and
$\Gamma|\vec{e_2}|$. The substitution of expressions (51), (52) into
Eqs. (47) and (48) lead to the Landau-Silin kinetic equations with
the precision to the terms of the indicated order. The same is
valid, if we consider the case $e_\downarrow=e_1+|\vec{e_2}|$. Thus,
the kinetic equations of the phenomenological theory, (the Eqs.
(7.21) and (7.23) in \cite{22}) are valid in the case of finite
temperature up to the terms linear in $\Gamma$, provided $\Gamma$
does not exceed the spin splitting of the energy levels.

\end{document}